\documentclass[12pt]{article}
\usepackage{amsmath}
\usepackage{amssymb}
\usepackage{epsfig}
\begin{document}

\numberwithin{equation}{section}

\begin{table}[t]
  \begin{flushright}
     CBPF-NF-003/06    \\
  \end{flushright}
\end{table}

\title{Randall-Sundrum scenario from 
$ D \! = \! 5 $, $ \mathcal{N} \! = \! 2 $
gauged Yang-Mills/Einstein/tensor supergravity}

\author{Richard S. Garavuso\thanks{garavuso@cbpf.br} \\
\emph{\normalsize Centro Brasileiro de Pesquisas F\'{\i}sicas}
\\
\emph{\normalsize Rio de Janeiro, RJ 22290-180, Brazil}}

\date{}

\maketitle
\thispagestyle{empty}

\begin{abstract}

In this paper, a new locally supersymmetric two brane 
Randall-Sundrum model is constructed.  
The construction starts from a 
$ D \! = \! 5 $, $ \mathcal{N} \! = \! 2 $
gauged Yang-Mills/Einstein/tensor supergravity theory 
with scalar manifold
$ \mathcal{M} = SO(1,1) \times SO(2,1) / SO(2) $ 
and gauge group
$ U(1)_{R} \times SO(2) $.
Here, $ U(1)_{R} $ is a subgroup of the $ R $-symmetry group $ SU(2)_{R} $
and $ SO(2) $ is a subgroup of the isometry group of $ \mathcal{M} $.
Next, the $ U(1)_{R} $ gauge coupling $ g_{R} $ is replaced by 
$ g_{R} \, \textrm{sgn}(x^{5}) $ and the fifth dimension is compactified 
on $ S^{1} / \mathbb{Z}_{2} $.
The conditions of local supersymmetry for the bulk plus brane system admit 
a Randall-Sundrum vacuum solution with constant scalars.  
This vacuum preserves 
$ \mathcal{N} \! = \! 2 $ supersymmetry in the $ AdS_{5} $ bulk and 
$ \mathcal{N} \! = \! 1 $ supersymmetry on the Minkowski 3-branes.

\end{abstract}

\pagebreak

\section{Introduction}

The two brane Randall-Sundrum scenario provides a mechanism for solving 
the hierarchy problem.
To see this, consider the original Randall-Sundrum model 
\cite{RandSund:ALarge} in which five dimensional pure anti-de Sitter gravity 
is compactified on an orbifold 
$ S^{1} / \mathbb{Z}_{2} $.\footnote{\label{spacetime}The $ D=5 $ spacetime
manifold has coordinates
$ x^{ \tilde{\mu} } = (x^{\mu}, x^{5}) $,
f\"{u}nfbein
$ e^{ \tilde{m} }_{ \tilde{\mu} } $,
and metric
$ g_{ \tilde{\mu} \tilde{\nu} }
  = e^{ \tilde{m} }_{ \tilde{\mu} } e^{ \tilde{n} }_{ \tilde{\nu} }
    \eta_{ \tilde{m} \tilde{n} } $.
Here
$ \eta_{ \tilde{m} \tilde{n} }
  = \textrm{diag}(-1,1,1,1,1)_{ \tilde{m} \tilde{n} } $,
$ \tilde{m}, \tilde{n}, \tilde{p}, \ldots
  = \dot{0}, \dot{1}, \dot{2}, \dot{3}, \dot{5} $,
and
$ \tilde{\mu}, \tilde{\nu}, \tilde{\rho}, \ldots = 0,1,2,3,5 $.
$ S^{1} / \mathbb{Z}_{2} $ is tangent to $ x^{5} $ with
$ \mathbb{Z}_{2} $ acting as $ x^{5} \rightarrow - x^{5} $.
In the \emph{upstairs} picture the orbifold is a circle
$ S^{1} $ of radius $ \rho $ with 
$ - \pi \rho \leq x^{5} \leq \pi \rho $.
Thus, there exist two hyperplanes (3-branes),
one at $ x^{5} = 0 $ and the other at $ x^{5} = \pi \rho $,
which are fixed under the $ \mathbb{Z}_{2} $ action.
In the \emph{downstairs} picture the orbifold is an interval
$ 0 \leq x^{5} \leq \pi \rho $
with the 3-branes forming boundaries to the spacetime manifold.}
There are two 3-branes, one at each orbifold fixed point,
with nonzero tension.
The 3-brane at 
$ x^{5} = 0 $ has tension $ \mathcal{T}^{(0)} $ 
and the 3-brane at 
$ x^{5} = \pi \rho $ has tension $ \mathcal{T}^{(\pi \rho)} $.
These 3-branes may support $ (3+1) $ dimensional field theories.
The minimal action is
\begin{align}
\label{Smin}   
S &= S_{bulk} + S_{brane}
\\
\label{Sbulk}
S_{bulk}  &= \int d^{5}x \, e 
             \left( - \frac{1}{2} M^{3} \mathcal{R}  - \Lambda \right) 
\\
\label{Sbrane}
S_{brane} &= - \int d^{5}x \, e^{(4)} 
             \left[
                     \mathcal{T}^{(0)} \delta( x^{5} )   
                   + \mathcal{T}^{ (\pi \rho) } \delta( x^{5} - \pi \rho )
             \right]
\end{align}   
where $ M $, $ \mathcal{R} $, and $ \Lambda $ are respectively the 
five dimensional Planck mass scale, 
five dimensional Ricci scalar, 
and bulk cosmological constant,  
$ e \equiv \det{  ( e^{ \tilde{m} }_{ \tilde{\mu} } )  } $, and 
$ e = e^{(4)} e^{ \dot{5} }_{5} $ on a brane. 
A solution to the five dimensional vacuum Einstein equations
\begin{equation}
\mathcal{R}_{ \tilde{\mu} \tilde{\nu} }
   - \frac{1}{2} g_{ \tilde{\mu} \tilde{\nu} } \mathcal{R}
   = - M^{-3} 
       \left\{
                 g_{ \tilde{\mu} \tilde{\nu} } \Lambda
              +  g_{ \tilde{\mu} \tilde{\nu} } 
                 \delta^{\mu}_{ \tilde{\mu} } \delta^{\nu}_{ \tilde{\nu} }
                 \frac{ e^{(4)} }{e} 
                 \left[ 
                      \mathcal{T}^{(0)} \delta( x^{5} )
                   +  \mathcal{T}^{(\pi \rho)} \delta( x^{5} - \pi \rho )
                 \right]
       \right\}
\end{equation}
which preserves four dimensional Poincar\'{e} invariance is given by 
\begin{equation}
\label{LineElement}
(ds)^{2} = a^{2}(x^{5}) \eta_{\mu \nu} dx^{\mu} dx^{\nu} + (dx^{5})^{2}
\end{equation}
where
\begin{equation}
a(x^{5}) = e^{ -k |x^{5}| }
\label{WarpFactor}
\end{equation}
is the \emph{warp factor} and $ \frac{1}{k} $ is the $ AdS_{5} $ curvature
radius.  This solution is valid provided 
$ \mathcal{T}^{(0)} $, $ \mathcal{T}^{(\pi \rho)} $, and $ \Lambda $ 
are related by
\begin{equation}
\label{FineTuning}
\mathcal{T}^{(0)} = - \mathcal{T}^{(\pi \rho)} 
  = - \frac{ \Lambda }{k}  = 6 M^{3} k. 
\end{equation}
The exponential warp factor can generate a large hierarchy of scales.  The 
four dimensional reduced Planck scale 
$ M_{P} $ can be expressed in terms of
$ M $ as follows
\begin{equation}
M^{2}_{P} = M^{3} \int_{-\pi \rho}^{\pi \rho} dx^{5} \, a^{2}(x^{5})
          = \frac{ M^{3} }{k} \left ( 1 - e^{ -2 \pi k \rho } \right). 
\end{equation}
The effective mass scales on the 3-branes at 
$ x^{5} = 0 $ and $ x^{5} = \pi \rho $ are respectively
$ M_{P} $ and $ M_{P} e^{ - \pi k \rho } $. 
If the Standard Model fields live on the 3-brane at 
$ x^{5} = \pi \rho $, then $ M_{P} e^{ - \pi k \rho } $ may be 
associated with the electroweak scale.

Locally supersymmetric two brane Randall-Sundrum models 
can be constructed from 
$ D \! = \! 5 $, $ \mathcal{N} \! = \! 2 $ 
gauged supergravity \cite{CerDal:General} theories.
The first such models
\cite{AltBagNem:Super,GhePom:Bulk,FalLalPok:SuperBranes}
were constructed from
$ D \! = \! 5 $, $ \mathcal{N} \! = \! 2 $
gauged pure supergravity \cite{gauged_pure} theories.
Subsequent examples include a model \cite{BerKalVan:SuperInSing}
constructed from a
$ D \! = \! 5 $, $ \mathcal{N} \! = \! 2 $
gauged Maxwell/Einstein supergravity 
\cite{GunSieTow:GaugingM/E,GunSieTow:Vanishing} theory
and models \cite{FalLalPok:Five} constructed from various
$ D \! = \! 5 $, $ \mathcal{N} \! = \! 2 $
gauged supergravity theories with hypermultiplets.

In this paper, a new locally supersymmetric two brane Randall-Sundrum 
model is constructed.
The construction starts from a
$ D \! = \! 5 $, $ \mathcal{N} \! = \! 2 $
gauged Yang-Mills/Einstein/tensor supergravity 
\cite{tensor,tensorAbelian,diag} theory with scalar manifold
$ \mathcal{M} = SO(1,1) \times SO(2,1) / SO(2) $ and 
$ U(1)_{R} \times SO(2) $ gauge group \cite{tensorAbelian}.
Next, the $ U(1)_{R} $ gauge coupling $ g_{R} $ is replaced by
$ g_{R} \, \textrm{sgn}(x^{5}) $ and the fifth dimension is compactified 
on $ S^{1} / \mathbb{Z}_{2} $.\footnote{The signum function
$ \textrm{sgn}(x^{5}) $ is $ +1 $ for $ 0 < x^{5} < \pi \rho $ and
$ -1 $ for $ -\pi \rho < x^{5} < 0 $.  It obeys
$ \partial_{5} \, \textrm{sgn}(x^{5})
  = 2 [ \delta(x^{5}) -  \delta(x^{5} - \pi \rho)] $.}
The modification of $ g_{R} $ allows a 
$ \mathbb{Z}_{2} $ invariant bulk theory to be constructed.
However, due to the presence of the signum function, the supersymmetry 
variation of the bulk action vanishes everywhere except at the 
$ \mathbb{Z}_{2} $ fixed points.  
The variation of the brane action cancels that of the bulk when the relations
(\ref{FineTuning}) are satisfied.
In these circumstances, the vacuum solution is given by 
(\ref{LineElement}) with warp factor (\ref{WarpFactor}).
This vacuum solution preserves $ \mathcal{N} \! = \! 2 $ supersymmetry in 
the bulk and $ \mathcal{N} \! = \! 1 $ supersymmetry on the branes.

This paper is organized as follows: 
The `starting point'
$ D \! = \! 5 $, $ \mathcal{N} \! = \! 2 $
gauged Yang-Mills/Einstein/tensor supergravity theory with gauge group
$ U(1)_{R} \times SO(2) $ is summarized in Section \ref{Review}.
In Section \ref{Orbifold}, a locally supersymmetric two brane 
Randall-Sundrum model is constructed from the starting point theory.
The results are discussed in Section \ref{Discussion}.

\section{\label{Review}Starting point theory}

This section summarizes the
$ D \! = \! 5 $, $ \mathcal{N} \! = \! 2 $
gauged Yang-Mills/Einstein/tensor supergravity 
\cite{tensor,tensorAbelian,diag} theory with
$ U(1)_{R} \times SO(2) $ gauge group \cite{tensorAbelian} which is taken 
as the starting point of the construction in Section \ref{Orbifold}.
The theory describes 
$ D \! = \! 5 $, $ \mathcal{N} \! = \! 2 $ pure supergravity \cite{pure}
coupled to one vector multiplet and one self-dual tensor multiplet,
with gauging of:
\begin{itemize}

\item[(1)]
{
A $ U(1)_{R} $ subgroup of the $ R $-symmetry group $ SU(2)_{R} $.
}

\item[(2)] 
{
An $ SO(2) $ subgroup of the isometry group
$ G = SO(2,1) \times SO(1,1) $
of the scalar field target manifold
$ \mathcal{M} = SO(1,1) \times SO(2,1) / SO(2) $.\footnote
{
\label{Jordan}
The \emph{generic Jordan family} of 
$ D \! = \! 5 $, $ \mathcal{N} \! = \! 2 $
gauged Yang-Mills/Einstein/tensor supergravity theories 
is characterized by scalar manifolds of the form
$ \mathcal{M} = SO(1,1) \times SO(n + 2m - 1,1) / SO(n + 2m - 1) $,
where $ n $ and $ m $ are respectively the number of vector and
self-dual tensor multiplets \cite{tensorAbelian}.
}
}

\end{itemize}

The fields of the \emph{pure supergravity} multiplet are the f\"{u}nfbein
$ e^{ \tilde{m} }_{ \tilde{\mu} } $,
an $ SU(2)_{R} $ doublet of symplectic Majorana gravitini
$ \Psi^{i}_{ \tilde{\mu} } $,
and the  graviphoton vector field  $ A^{0}_{ \tilde{\mu} } $.\footnote{
The $ SU(2)_{R} $ indices 
$ i,j,k,\ldots = 1,2 $ are raised and lowered according to
$ X^{i} = \epsilon^{ij} X_{j} $,
$ X_{i} =  X^{j} \epsilon_{ji} $ with
$ \epsilon^{ij} $ and $ \epsilon_{ij} $ antisymmetric and
$ \epsilon^{12} = \epsilon_{12} = 1 $.
}
The \emph{vector} multiplet contains
a vector field $ A^{1}_{ \tilde{\mu} } $,
an $ SU(2)_{R} $ doublet of symplectic Majorana spin-1/2 fermions
$ \lambda^{i1} $, and
a real scalar field $ \phi^{1} $.
Finally, the \emph{self-dual tensor} multiplet contains
a self-dual antisymmetric tensor field with real and imaginary parts
$ B^{2}_{ \tilde{\mu} \tilde{\nu} } $ and
$ B^{3}_{ \tilde{\mu} \tilde{\nu} } $, respectively,
two $ SU(2)_{R} $ doublets of symplectic Majorana spin-1/2 fermions
$ \lambda^{i2} $ and $ \lambda^{i3} $,
and two real scalar fields
$ \phi^{2} $ and $ \phi^{3} $. Thus, the total field content is
\begin{equation}
\{ e^{ \tilde{m} }_{ \tilde{\mu} }, \Psi^{i}_{ \tilde\mu },
A^{I}_{\tilde{\mu}}, B^{M}_{ \tilde{\mu} \tilde{\nu} },
\lambda^{ i \tilde{a} }, \phi^{ \tilde{x} } \}
\end{equation}
where 
\begin{align*}
I,J,K,\ldots &= 0,1,
\quad
M,N,P,\ldots = 2,3,
\\
\tilde{a},\tilde{b},\tilde{c},\ldots &= 1,2,3,
\quad
\tilde{x},\tilde{y},\tilde{z},\ldots = 1,2,3.
\end{align*}
It is convenient for notational purposes to define 
$ \tilde{I} \equiv (I,M) $.

The dreibein, metric, and $ SO(3) $ spin connection on the manifold
$ \mathcal{M} $ are denoted by
$ f^{ \tilde{a} }_{ \tilde{x} } $,
$ g_{ \tilde{x} \tilde{y} } $, and
$ \Omega^{ \tilde{a} \tilde{b} }_{ \tilde{x} } $, respectively.
This manifold is embedded \cite{MESG} in a four dimensional ambient space 
with coordinates 
$ \xi^{ \tilde{I} }( \phi^{ \tilde{x} }, N) $,
cubic norm
$ N(\xi) \! = \! \left( \frac{2}{3} \right)^{ \frac{3}{2} } 
                 C_{ \tilde{I} \tilde{J} \tilde{K} }
                 \xi^{ \tilde{I} } \xi^{\tilde{J} } \xi^{\tilde{K} } $,
and metric
$ a_{ \tilde{I} \tilde{J} } \, = \!
     - \frac{1}{2} \partial_{ \tilde{I} \tilde{J} } \ln{N(\xi)} $.
$ \mathcal{M} $ corresponds to the $ N(\xi) = 1 $ hypersurface
\begin{equation}
C_{ \tilde{I} \tilde{J} \tilde{K} }
h^{ \tilde{I} } h^{\tilde{J} } h^{\tilde{K} } = 1
\quad
\textrm{with}
\quad
h^{\tilde{I}} \equiv \textstyle{\sqrt{\frac{2}{3}}} 
\xi^{\tilde{I}}|_{N=1}.
\end{equation}
A basis \cite{tensorAbelian} can be chosen in which the nonvanishing 
$ C_{ \tilde{I} \tilde{J} \tilde{K} } $ are
\begin{equation}
C_{0\tilde{I}\tilde{I}} = C_{\tilde{I}0\tilde{I}} = 
C_{\tilde{I}\tilde{I}0} 
\quad
\textrm{with}
\quad
C_{011} = - C_{022} = -C_{033} = \frac{ \sqrt{3} }{2} .
\end{equation}
In this basis, the constraint $ N(\xi) \! = \! 1 $ is solved by
\begin{gather}
h^{0} = \frac{1}{ \sqrt{3} \| \phi \|^{2} },
\quad
h^{1} = \sqrt{ \frac{2}{3} } \phi^{1},
\quad
h^{2} = \sqrt{ \frac{2}{3} } \phi^{2},
\quad
h^{3} = \sqrt{ \frac{2}{3} } \phi^{3}
\nonumber
\\
\textrm{with}
\quad
\| \phi \|^{2}  
  \equiv (\phi^{1})^{2} - (\phi^{2})^{2} - (\phi^{3})^{2}.
\end{gather}
Lowering the index of $ h^{\tilde{I}} $ with 
$ \stackrel{\circ}{a}_{ \tilde{I} \tilde{J} } \, \equiv
                    a_{ \tilde{I} \tilde{J} }|_{N=1} $
yields
$ h_{\tilde{I}} \equiv
     \frac{1}{ \sqrt{6} } \partial_{ \tilde{I} } N(\xi)|_{N=1} $.

The fermions 
$ \Psi^{i}_{ \tilde{\mu} } $ and 
$ \lambda^{ i \tilde{a} } $
are $ U(1)_{R} $ charged, whereas the fields
$ \phi^{ \tilde{x} } $,
$ \lambda^{ i \tilde{a} } $, and
$ B^{M}_{ \tilde{\mu} \tilde{\nu} } $ 
carry charge under $ SO(2) $.
Denoting the $ U(1)_{R} $ and $ SO(2) $ couplings by 
$ g_{R} $ and $ g $, respectively, the 
$ U(1)_{R} \times SO(2) $ gauge covariant derivatives of these fields are
\begin{align}
\mathfrak{D}_{ \tilde{\mu} } \Psi^{i}_{ \tilde{\nu} }
   &\equiv   \nabla_{ \tilde{\mu} } \Psi^{i}_{ \tilde{\nu} }
           + g_{R} A^{I}_{ \tilde{\mu} } \mathcal{P}^{i}_{Ij} 
             \Psi^{j}_{ \tilde{\nu} }
\\
\mathfrak{D}_{ \tilde{\mu} } \lambda^{ i \tilde{a} }
   &\equiv   \nabla_{ \tilde{\mu} } \lambda^{ i \tilde{a} }
           + g_{R} A^{I}_{ \tilde{\mu} } \mathcal{P}^{i}_{Ij}
             \lambda^{ j \tilde{a} }
           + g A^{I}_{ \tilde{\mu} } L^{ \tilde{a} }_{ I \tilde{b} } 
             \lambda^{ i \tilde{b} }
\\
\mathfrak{D}_{ \tilde{\mu} } \phi^{ \tilde{x} }
   &\equiv   \partial_{ \tilde{\mu} } \phi^{ \tilde{x} } 
           + g A^{I}_{ \tilde{\mu} } K^{ \tilde{x} }_{I}
\\
\mathfrak{D}_{ \tilde{\mu} } B^{M}_{ \tilde{\nu} \tilde{\rho} }
   &\equiv   \nabla_{ \tilde{\mu} } B^{M}_{ \tilde{\nu} \tilde{\rho} }
           + g A^{I}_{ \tilde{\mu} } \Lambda^{M}_{IN} 
             B^{N}_{ \tilde{\nu} \tilde{\rho} }.     
\end{align}
Here, $ \nabla_{ \tilde{\mu} } $ is the spacetime covariant derivative.  
$ K^{ \tilde{x} }_{I} $ are the Killing vectors on 
$ \mathcal{M} $ that generate $ SO(2) $.  
$ L^{ \tilde{a} }_{I \tilde{b} } 
   \equiv \partial_{ \tilde{b} } K^{ \tilde{a} }_{I} $ and
$ \Lambda^{M}_{IN} \! = \! \frac{2}{\sqrt{6}} \Omega^{MP} C_{IPN} $
are the $ SO(2) $ transformation matrices of 
$ \lambda^{ i \tilde{a} } $ and
$ B^{M}_{ \tilde{\nu} \tilde{\rho} } $, respectively.\footnote{
$ \Omega_{MN} \Omega^{NP} = \delta^{P}_{M} $,
$ \Omega_{MN} = - \Omega_{NM} $, and
$ \Omega^{23} = - \Omega^{32} = -1 $.
}
The $ SU(2)_{R} $ valued prepotentials $ \mathcal{P}^{i}_{Ij} $ are chosen 
along the $ \sigma_{3} $ direction,\footnote{Note that in 
\cite{tensorAbelian}, the $ \sigma_{2} $ direction is chosen.  Gaugings
of different $ U(1)_{R} $ subgroups of $ SU(2)_{R} $ can be rotated into
each other and hence are physically equivalent.} i.e.
\begin{equation}
\label{prepotential}
\mathcal{P}^{i}_{Ij} = V_{I} Q^{i}{}_{j}
\quad
\textrm{with}
\quad
Q^{i}{}_{j} \equiv i (\sigma_{3})^{i}{}_{j}.
\end{equation}
The $ V_{I} $ are real constants that define the linear combination of the 
vector fields $ A^{I}_{ \tilde{\mu} } $ that is used as the $ U(1)_{R} $
gauge field
\begin{equation}
A_{ \tilde{\mu} } \left[ U(1)_{R} \right] = V_{I} A^{I}_{ \tilde{\mu} }.
\end{equation}
The Abelian field strengths
$ F^{I}_{ \tilde{\mu} \tilde{\nu} } 
   =   \partial_{ \tilde{\mu} } A^{I}_{ \tilde{\nu} }
     - \partial_{ \tilde{\nu} } A^{I}_{ \tilde{\mu} } $   
and the $ B^{M}_{ \tilde{\mu} \tilde{\nu} } $ are combined to form
$ H^{ \tilde{I} }_{ \tilde{\mu} \tilde{\nu} } 
  \equiv \left( F^{I}_{ \tilde{\mu} \tilde{\nu} },
                B^{M}_{ \tilde{\mu} \tilde{\nu} } \right) $.

The Lagrangian is (up to 4-fermion terms) \cite{tensor}
\begin{align}
e^{-1} \mathcal{L} =
  &- \frac{1}{2} \mathcal{R}
   - \frac{1}{2} \bar{\Psi}^{i}_{ \tilde{\mu} }
     \Gamma^{ \tilde{\mu} \tilde{\nu} \tilde{\rho} }
     \mathfrak{D}_{\tilde{\nu}}
     \Psi_{ \tilde{\rho } i}
   - \frac{1}{4} \stackrel{\circ}{a}_{ \tilde{I} \tilde{J} }
     H^{ \tilde{I} }_{ \tilde{\mu} \tilde{\nu} }
     H^{ \tilde{J} \tilde{\mu} \tilde{\nu} }
\nonumber
\displaybreak[0]\\
  &- \frac{1}{2} \bar{\lambda}^{i}_{ \tilde{a} }
     \left(  \Gamma^{ \tilde{\mu} } \mathfrak{D}_{ \tilde{\mu} }
             \delta^{ \tilde{a} \tilde{b} }
           + \Omega^{ \tilde{a} \tilde{b} }_{ \tilde{x} }
             \Gamma^{ \tilde{\mu} } \mathfrak{D}_{ \tilde{\mu} }
             \phi^{ \tilde{x} }
     \right) \lambda_{ i \tilde{b} } 
   - \frac{1}{2} g_{ \tilde{x} \tilde{y} }
     ( \mathfrak{D}_{ \tilde{\mu} } \phi^{ \tilde{x} } )
     ( \mathfrak{D}^{ \tilde{\mu} } \phi^{ \tilde{y} } ) 
\nonumber
\displaybreak[0]\\
  &- \frac{i}{2} \bar{\lambda}^{i}_{ \tilde{a} } 
     \Gamma^{ \tilde{\mu} } \Gamma^{ \tilde{\nu} }
     \Psi_{ \tilde{\mu} i } f^{ \tilde{a} }_{ \tilde{x} }
     \mathfrak{D}_{ \tilde{\nu} } \phi^{ \tilde{x} }
   + \frac{1}{4} h^{ \tilde{a} }_{ \tilde{I} }
     \bar{\lambda}^{i}_{ \tilde{a} }
     \Gamma^{ \tilde{\mu} } \Gamma^{ \tilde{\lambda} \tilde{\rho} }
     \Psi_{ \tilde{\mu} i }
     H^{ \tilde{I} }_{ \tilde{\lambda} \tilde{\rho} }
\nonumber
\displaybreak[0]\\
  &+ \frac{i}{ 2 \sqrt{6} }
     \left(  \frac{1}{4} \delta_{ \tilde{a} \tilde{b} } h_{ \tilde{I} }
           + T_{ \tilde{a} \tilde{b} \tilde{c} }
             h_{ \tilde{I} }^{ \tilde{c} }
     \right)
     \bar{\lambda}^{ i \tilde{a} }
     \Gamma^{ \tilde{\mu} \tilde{\nu} } \lambda^{ \tilde{b} }_{i}
     H^{ \tilde{I} }_{ \tilde{\mu} \tilde{\nu} }
\nonumber
\displaybreak[0]\\
  &- \frac{3i}{ 8\sqrt{6} } h_{ \tilde{I} }
     \left( \bar{\Psi}^{i}_{\tilde{\mu}}
     \Gamma^{ \tilde{\mu} \tilde{\nu} \tilde{\rho} \tilde{\sigma} }
    \Psi_{ \tilde{\nu} i }
     H^{ \tilde{I} }_{ \tilde{\rho} \tilde{\sigma} }
   + 2 \bar{\Psi}^{ \tilde{\mu} i } \bar{\Psi}^{\tilde{\nu}}_{i}
     H^{ \tilde{I} }_{ \tilde{\mu} \tilde{\nu} } \right)
\nonumber
\displaybreak[0]\\
  &+ \frac{e^{-1}}{ 6\sqrt{6} } C_{IJK}
     \epsilon^{ \tilde{\mu} \tilde{\nu} \tilde{\rho} \tilde{\sigma}
                \tilde{\lambda} }
        F^{I}_{ \tilde{\mu} \tilde{\nu} }
        F^{J}_{ \tilde{\rho} \tilde{\sigma} }
        A^{K}_{ \tilde{\lambda} }
  + \frac{ e^{-1} }{4g} 
    \epsilon^{ \tilde{\mu} \tilde{\nu} \tilde{\rho} 
                  \tilde{\sigma} \tilde{\lambda} }
    \Omega_{MN} B^{M}_{ \tilde{\mu} \tilde{\nu} }
    \mathfrak{D}_{ \tilde{\rho} } B^{N}_{ \tilde{\sigma} \tilde{\lambda} }
\nonumber     
\displaybreak[0]\\
 &+ g \bar{\lambda}^{i}_{ \tilde{a} } \Gamma^{ \tilde{\mu} }
      \Psi_{ \tilde{\mu} i } W^{ \tilde{a} }
  + g \bar{\lambda}^{i}_{ \tilde{a} } \lambda_{i \tilde{b} } 
      W^{ \tilde{a} \tilde{b} } 
\nonumber
\displaybreak[0]\\
  &+ \frac{ i \sqrt{6} }{4} g_{R} 
     \bar{\Psi}^{i}_{ \tilde{\mu} }
     \Gamma^{ \tilde{\mu} \tilde{\nu} } \Psi^{j}_{ \tilde{\nu} }
     \mathcal{P}_{ij}
   + g_{R} \bar{\lambda}^{ i \tilde{a} }
     \Gamma^{ \tilde{\mu} } \Psi^{j}_{ \tilde{\mu} }
     \mathcal{P}_{ \tilde{a} ij }
   - \frac{i}{ 2 \sqrt{6} } g_{R} \bar{\lambda}^{ i \tilde{a} }
       \lambda^{ j \tilde{b} } \mathcal{P}_{ \tilde{a} \tilde{b} ij }
\nonumber
\displaybreak[0]\\
  &- g^{2} P - g^{2}_{R} P^{(R)}
\label{L_{GYME}}
\end{align}
and the transformation laws are (to leading order in fermion fields)
\begin{align}
\delta e^{ \tilde{m} }_{ \tilde{\mu} }
   &=  \frac{1}{2} \bar{\varepsilon}^{i} \Gamma^{\tilde{m}}
       \Psi_{ \tilde{\mu} i }
\displaybreak[0]\\
\label{deltaPsi^{i}}
\delta \Psi_{ \tilde{\mu} i }
   &=   \mathfrak{D}_{ \tilde{\mu} } \varepsilon_{i}
      + \frac{i}{ 4 \sqrt{6} } h_{ \tilde{I} }
        (   \Gamma_{ \tilde{\mu} }{}^{ \tilde{\nu} \tilde{\rho} }
          - 4 \delta^{ \tilde{\nu} }_{ \tilde{\mu} } 
            \Gamma^{ \tilde{\rho} }
        )
        H^{ \tilde{I} }_{\tilde{\nu} \tilde{\rho} } 
        \varepsilon_{i}
      + \frac{i}{\sqrt{6} } g_{R} \Gamma_{ \tilde{\mu} }
        \mathcal{P}_{ij} \varepsilon^{j}
\displaybreak[0]\\
\delta A^{I}_{ \tilde{\mu} }
   &= \vartheta^{I}_{ \tilde{\mu} }
\displaybreak[0]\\
\delta B^{M}_{ \tilde{\mu} \tilde{\nu} }
   &=   2 \mathfrak{D}_{ [ \tilde{\mu} } \vartheta^{M}_{ \tilde{\nu} ] }
      + \frac{ \sqrt{6} g }{4} \Omega^{MN} h_{N}
        \bar{\Psi}^{i}_{ [ \tilde{\mu} } \Gamma_{ \tilde{\nu} ] } 
        \varepsilon_{i}
      + \frac{ig}{4} \Omega^{MN} h_{ N \tilde{a} }
        \bar{\lambda}^{ i \tilde{a} } 
        \Gamma_{ \tilde{\mu} \tilde{\nu} } \varepsilon_{i}
\displaybreak[0]\\
\delta \lambda^{ \tilde{a} }_{i}
   &= - \frac{i}{2} f^{ \tilde{a} }_{ \tilde{x} } \Gamma^{ \tilde{\mu} }
        ( \mathfrak{D}_{ \tilde{\mu} } \phi^{ \tilde{x} } ) 
        \varepsilon_{i}
      + \frac{1}{4} h^{ \tilde{a} }_{ \tilde{I} }
        \Gamma^{ \tilde{\mu} \tilde{\nu} } \varepsilon_{i}
        H^{ \tilde{I} }_{ \tilde{\mu} \tilde{\nu} }
      + g W^{ \tilde{a} } \varepsilon_{i}
      + g_{R} \mathcal{P}^{ \tilde{a} }_{ij} \varepsilon^{j}
\\
\delta \phi^{ \tilde{x} }
   &= \frac{i}{2} f^{ \tilde{x} }_{ \tilde{a} }
     \bar{\varepsilon}^{i} \lambda^{ \tilde{a} }_{i}
\end{align}
with
\begin{equation}
\vartheta^{ \tilde{I} }_{ \tilde{\mu} }
  \equiv - \frac{1}{2} h^{ \tilde{I} }_{ \tilde{a} } \bar{\varepsilon}^{i}
           \Gamma_{ \tilde{\mu} } \lambda^{ \tilde{a} }_{i}
         + \frac{i \sqrt{6} }{4} h^{ \tilde{I} } 
           \bar{\Psi}^{i}_{ \tilde{\mu} }\varepsilon_{i}.
\end{equation}
The scalar field dependent quantities
$ h_{ \tilde{I} } $, $ h^{ \tilde{I} } $, $ h^{\tilde{a}}_{ \tilde{I} } $, 
$ h^{ \tilde{I} }_{ \tilde{a} } $, $ T_{ \tilde{a}\tilde{b}\tilde{c} } $, 
and $ \stackrel{\circ}{a}_{ \tilde{I} \tilde{J} } $ 
are subject to the following constraints imposed by supersymmetry
\cite{MESG}:
\begin{align*}
C_{\tilde{I}\tilde{J}\tilde{K}} &=  
    \frac{5}{2} h_{\tilde{I}} h_{\tilde{J}} h_{\tilde{K}}
  - \frac{3}{2} \stackrel{\circ}{a}_{(\tilde{I}\tilde{J}}h_{\tilde{K})}
  + T_{\tilde{x}\tilde{y}\tilde{z}} h^{\tilde{x}}_{\tilde{I}} 
    h^{\tilde{y}}_{\tilde{J}} h^{\tilde{z}}_{\tilde{K}},
\\
h^{\tilde{I}} h_{\tilde{I}} &= 1,
\quad
h^{\tilde{I}}_{\tilde{x}} h_{\tilde{I}} 
  = h_{\tilde{I}\tilde{x}} h^{\tilde{I}} = 0,
\quad
h^{\tilde{I}}_{\tilde{x}} h^{\tilde{J}}_{\tilde{y}} 
\stackrel{\circ}{a}_{\tilde{I}\tilde{J}} \, = g_{\tilde{x}\tilde{y}},
\\
\stackrel{\circ}{a}_{\tilde{I}\tilde{J}} &= 
   h_{\tilde{I}} h_{\tilde{J}} + h^{\tilde{x}}_{\tilde{I}} 
   h^{\tilde{y}}_{\tilde{J}} g_{\tilde{x}\tilde{y}},
\\
h_{\tilde{I},\tilde{x}} &= 
   \sqrt{ \textstyle{ \frac{2}{3} } } h_{\tilde{I}\tilde{x}},
\quad
h^{\tilde{I}}{}_{,\tilde{x}} = 
  - \sqrt{ \textstyle{ \frac{2}{3} } } h^{\tilde{I}}_{\tilde{x}},
\\
h_{\tilde{I}\tilde{x};\tilde{y}} &= 
   \sqrt{ \textstyle{ \frac{2}{3} } }
   \left( g_{\tilde{x}\tilde{y}} h_{\tilde{I}} 
 + T_{\tilde{x}\tilde{y}\tilde{z}} h^{\tilde{z}}_{\tilde{I}} \right),
\quad
h^{\tilde{I}}_{\tilde{x};\tilde{y}} 
 = - \sqrt{ \textstyle{ \frac{2}{3} } }
     \left(g_{\tilde{x}\tilde{y}} h^{\tilde{I}} 
   + T_{\tilde{x}\tilde{y}\tilde{z}} h^{\tilde{I}\tilde{z}} \right).
\end{align*}

The presence of the tensor fields introduces the terms proportional to
\begin{align}
W^{ \tilde{a} }
  &= - \frac{ \sqrt{6} }{8} h^{ \tilde{a} }_{M} \Omega^{MN} h_{N} 
\\
W^{ \tilde{a} \tilde{b} }
  &=   i h^{ J [ \tilde{a} } K^{ \tilde{b} ] }_{J}
     + \frac{ i \sqrt{6} }{4} h^{J} 
       K^{ [ \tilde{a} ; \tilde{b} ] }_{J}
\end{align}
and leads to the scalar potential contribution
\begin{equation}
g^{2} P = 2 W_{ \tilde{a} } W^{ \tilde{a} }.
\end{equation}
The gauging of $ U(1)_{R} $ introduces the terms proportional to
\begin{align}
\mathcal{P}_{ij} &\equiv h^{I} \mathcal{P}_{Iij} = h^{I} V_{I} Q_{ij}
\\   
\mathcal{P}_{\tilde{a}ij} 
  &\equiv h^{I}_{ \tilde{a} } \mathcal{P}_{Iij} 
   = h^{I}_{ \tilde{a} } V_{I} Q_{ij}
\\
\mathcal{P}_{ \tilde{a} \tilde{b} ij } 
  &\equiv   \delta_{ \tilde{a} \tilde{b} } \mathcal{P}_{ij} 
          + 4 T_{ \tilde{a} \tilde{b} \tilde{c} } 
            \mathcal{P}^{ \tilde{c} }_{ij}
\end{align}
and leads to the scalar potential contribution
\begin{align}
g^{2}_{R} P^{(R)}
  &= g^{2}_{R}
     \left(
       - 2 \mathcal{P}_{ij} \mathcal{P}^{ij}   
       + \mathcal{P}_{ \tilde{a} ij } \mathcal{P}^{ \tilde{a} ij }
     \right)
\\
  &= g^{2}_{R}
     ( 
       - 6 \mathcal{W}^{2} 
        + \frac{9}{2} 
          \mathcal{W}_{ , \tilde{a} } \mathcal{W}^{ , \tilde{a} }
     ) 
\quad
\textrm{where}
\quad
\mathcal{W} \equiv \textstyle{ \sqrt{ \frac{2}{3} } } h^{I} V_{I}.
\end{align}
The total scalar potential is
\begin{equation}
\mathcal{V} = g^{2} P + g^{2}_{R} P^{(R)}.
\end{equation}

\section{\label{Orbifold}Construction of model}

In this section, a locally supersymmetric two brane Randall-Sundrum 
model is constructed.\footnote
{
A chiral basis is chosen for the Dirac matrices
\begin{equation*}
\Gamma^{ \tilde{m} }
  = \left( \Gamma^{m}, \Gamma^{ \dot{5} } \right)
  = \left(
           \begin{bmatrix}
              0                   &  -i\sigma^{m}   \\
             -i \bar{\sigma}^{m}  &   0
           \end{bmatrix},
           \begin{bmatrix}
             -1                 &   0            \\
              0                 &   1
           \end{bmatrix}
    \right)   
\end{equation*}
where
$ \sigma^{m} = (1, \vec{\sigma} ) $
and
$ \bar{\sigma}^{m} = (1, - \vec{\sigma} ) $.
}
The starting point is the 
$ D \! = \! 5 $, $ \mathcal{N} \! = \! 2 $
gauged Yang-Mills/Einstein supergravity theory with gauge group
$ U(1)_{R} \times SO(2) $ which is summarized in Section \ref{Review}.
Next, the $ U(1)_{R} $ gauge coupling $ g_{R} $ is replaced by 
$ g_{R} \, \textrm{sgn}(x^{5}) $ and the fifth dimension is compactified 
on $ S^{1} / \mathbb{Z}_{2} $.\footnote
{
Note that the scalar potential 
$ \mathcal{V} = g^{2} P + g^{2}_{R} P^{(R)} $ 
is unchanged by
$ g_{R} \rightarrow g_{R} \, \textrm{sgn}(x^{5}) $ because
$ \textrm{sgn}^{2}(x^{5}) = 1 $.
} 
The modification of $ g_{R} $ allows a $ \mathbb{Z}_{2} $ invariant bulk 
theory to be constructed.  In this framework, the $ \mathbb{Z}_{2} $
symmetry does not commute with the $ SU(2)_{R} $ R-symmetry and gauging
different $ U(1)_{R} $ results in physically different theories.

The $ \mathbb{Z}_{2} $ action on each field is defined by boundary 
conditions at the $ \mathbb{Z}_{2} $ fixed points.
These boundary conditions must be chosen such that the Lagrangian is 
$ \mathbb{Z}_{2}  $ invariant.
A consistent set of boundary conditions is obtained as follows.
Under reflections $ x^{5} \rightarrow - x^{5} $ about the 
$ \mathbb{Z}_{2} $ fixed point $ x^{5} = 0 $, choose the bosonic fields
\begin{align*}
\Phi &= e^{m}_{\mu}, e^{ \dot{5} }_{5}, A^{I}_{5}, B^{M}_{ \mu 5 }, 
        \phi^{ \tilde{x} } 
\\
\Theta &= e^{\dot{5}}_{\mu}, e^{m}_{5}, A^{I}_{\mu}, B^{M}_{ \mu \nu }
\end{align*}
to satisfy
\begin{align}
\Phi( x^{\mu}, x^{5} ) &= + \Phi( x^{\mu}, -x^{5} )
\\
\Theta( x^{\mu}, x^{5} ) &= - \Theta( x^{\mu}, -x^{5} )
\end{align}
and choose the fermionic fields to satisfy
\begin{align}
\Gamma^{ \dot{5} }   \, \Psi^{i}_{\mu} ( x^{\mu},   x^{5} ) 
    &= +iQ^{i}{}_{j} \, \Psi^{j}_{\mu} ( x^{\mu}, - x^{5} )
\\
\Gamma^{ \dot{5} }   \, \Psi^{i}_{5} ( x^{\mu},   x^{5} ) 
    &= -iQ^{i}{}_{j} \, \Psi^{j}_{5} ( x^{\mu}, - x^{5} )
\\
\Gamma^{ \dot{5} }   \, \lambda^{ i \tilde{a} } ( x^{\mu},   x^{5} ) 
    &= -iQ^{i}{}_{j} \, \lambda^{ j \tilde{a} } ( x^{\mu}, - x^{5} )
\\
\label{BCone}
\Gamma^{ \dot{5} }   \, \varepsilon^{i} ( x^{\mu},   x^{5} ) 
    &= +iQ^{i}{}_{j} \, \varepsilon^{j} ( x^{\mu}, - x^{5} )
\end{align}
where $ Q^{i}{}_{j} $ is given by ($ \ref{prepotential} $).
Under reflections $ x^{5} \rightarrow - x^{5} $ about the
$ \mathbb{Z}_{2} $ fixed point $ x^{5} = \pi \rho $, 
choose the bosonic fields to satisfy
\begin{align}
\Phi( x^{\mu}, \pi \rho + x^{5} ) 
  &= + \Phi( x^{\mu}, \pi \rho -x^{5} )
\\
\Theta( x^{\mu}, \pi \rho + x^{5} ) 
  &= - \Theta( x^{\mu}, \pi \rho - x^{5} )
\end{align}
and choose the fermionic fields to satisfy
\begin{align}
\Gamma^{ \dot{5} } \, 
         \Psi^{i}_{\mu} ( x^{\mu}, \pi \rho + x^{5} )
    &=  +iQ^{i}{}_{j} \, \Psi^{j}_{\mu} ( x^{\mu}, \pi \rho - x^{5} )
\\
\Gamma^{ \dot{5} } \, 
         \Psi^{i}_{5} ( x^{\mu}, \pi \rho + x^{5} )  
    &=  -iQ^{i}{}_{j} \, 
         \Psi^{j}_{5} ( x^{\mu}, \pi \rho - x^{5} )   
\\
\Gamma^{ \dot{5} } \, 
         \lambda^{ i \tilde{a} } ( x^{\mu}, \pi \rho + x^{5} )
    &=  -iQ^{i}{}_{j} \, 
         \lambda^{ j \tilde{a} } ( x^{\mu}, \pi \rho - x^{5} )
\\
\label{BCtwo}
\Gamma^{ \dot{5} } \, 
         \varepsilon^{i} ( x^{\mu}, \pi \rho + x^{5} )
    &=  +iQ^{i}{}_{j} \, 
         \varepsilon^{j} ( x^{\mu}, \pi \rho - x^{5} ).
\end{align}

The bulk theory is now $ \mathbb{Z}_{2} $ invariant.
However, due to the presence of the signum function, the supersymmetry 
variation of the bulk Lagrangian vanishes everywhere except at the 
$ \mathbb{Z}_{2} $ fixed points:
\begin{align}
\label{BulkVariation}
\delta \mathcal{L}_{bulk}
 &=   \delta^{ (\Psi) } \mathcal{L}_{bulk}
    + \delta^{ (\lambda) } \mathcal{L}_{bulk}
\\
\delta^{ (\Psi) } \mathcal{L}_{bulk}
 &= - 6 g_{R} \mathcal{W}
      e^{(4)} \frac{1}{2}
      \bar{\Psi}^{i}_{ \mu } \Gamma^{ \mu } \Gamma^{ \dot{5} }
      (iQ_{ij}) \varepsilon^{j}
      \left[ \delta ( x^{5} ) - \delta ( x^{5} - \pi \rho ) \right]
\\
\delta^{ (\lambda) } \mathcal{L}_{bulk}
 &= - 6i g_{R} \mathcal{W}_{ , \tilde{x} }
      e^{(4)} \frac{1}{2}
      \bar{\lambda}^{ i \tilde{x} } \Gamma^{ \dot{5} }
      (iQ_{ij}) \varepsilon^{j}
      \left[ \delta ( x^{5} ) - \delta ( x^{5} - \pi \rho ) \right].
\end{align}
It follows from the boundary conditions (\ref{BCone}) 
and (\ref{BCtwo}) that
\begin{align}
\Gamma^{ \dot{5} } (iQ_{ij}) \varepsilon^{j} ( x^{\mu}, 0 )   
 &= \varepsilon_{i} ( x^{\mu}, 0 )
\\
\Gamma^{ \dot{5} } (iQ_{ij}) \varepsilon^{j} ( x^{\mu},\pi \rho )
 &= \varepsilon_{i} ( x^{\mu}, \pi \rho ).
\end{align}  
Thus,
\begin{align}
\delta^{ (\Psi) } \mathcal{L}_{bulk}
 &= - 6 g_{R} \mathcal{W}
      e^{(4)} \frac{1}{2}
      \bar{\Psi}^{i}_{ \mu } \Gamma^{ \mu } \varepsilon_{i}
      \left[
        \delta ( x^{5} ) - \delta ( x^{5} - \pi \rho )
      \right]
\\
\delta^{ (\lambda) } \mathcal{L}_{bulk}
 &= - 6i g_{R} \mathcal{W}_{ , \tilde{x} } 
      e^{(4)} \frac{1}{2}
      \bar{\lambda}^{ i \tilde{x} } \varepsilon_{i}
      \left[
        \delta ( x^{5} ) - \delta ( x^{5} - \pi \rho )
      \right].
\end{align}   

The bulk variation (\ref{BulkVariation}) 
can be cancelled by the variation of
\begin{gather}
\label{Lbrane}
\mathcal{L}_{brane} 
  = - e^{(4)}
      \left[
          \mathcal{T}^{ (0) } \delta( x^{5} )
        + \mathcal{T}^{ ( \pi \rho ) } \delta( x^{5} - \pi \rho )
      \right]
\nonumber
\\ 
\textrm{with}
\quad
\mathcal{T}^{(0)} = - \mathcal{T}^{(\pi \rho)} = 6 g_{R} \mathcal{W}.
\end{gather}
For supersymmetry to hold, the Killing spinor equations must be satisfied.
As demonstrated below, these equations admit the vacuum solution
\begin{equation}
\label{Ansatz}
(ds)^{2} = a^{2}(x^{5}) \eta_{\mu \nu} dx^{\mu} dx^{\nu} + (dx^{5})^{2}
\quad
\textrm{with}
\quad
a(x^{5}) = e^{ -k |x^{5}| }
\end{equation}
at a critical point $ \phi_{crit} $ of the scalar potential in which
\begin{equation}
\label{kDefined}
g_{R} \mathcal{W}_{crit} = \sqrt{ \frac{ - \Lambda }{6} } \equiv k.
\end{equation}
The scalar potential has such a critical point given by
\begin{equation}
\left( \phi^{1}_{crit} \right)^{3} = \sqrt{2} \frac{ V_{0} }{ V_{1} },
\quad
\phi^{2}_{crit} = \phi^{3}_{crit} = 0
\end{equation}
whenever $ V_{0} V_{1} > 0 $ \cite{tensorAbelian}.
The associated cosmological constant is
\begin{equation}
\Lambda \equiv \langle \mathcal{V} \rangle
  = - 6 g^{2}_{R} 
      \left( \phi^{1}_{crit} \right)^{2} \left( V_{1} \right)^{2}.
\end{equation}

Taking (\ref{Ansatz}) as an Ansatz,
the Killing spinor equations are\footnote{The only nonvanishing components 
of the spin connection 
$ \omega^{ \tilde{m} \tilde{n} }_{ \tilde{\mu} } $ are
$ \omega^{ m \dot{5} }_{ \mu } 
  = - \omega^{ \dot{5} m }_{ \mu }  
  = a^{\prime} \delta^{m}_{\mu} $,
where $ ' $ denotes partial differentiation with respect to $ x^{5} $.
Thus, 
$ \nabla_{\mu} \varepsilon_{i} 
    \equiv \left(  
              \partial_{\mu} 
            + \frac{1}{4} \omega^{ \tilde{m} \tilde{n} }_{ \mu }
              \Gamma_{ \tilde{m} \tilde{n} } 
           \right) \varepsilon_{i} 
     = \left[ 
           \partial_{ \mu } 
         + \frac{1}{2} \left( \frac{ a^{\prime} }{a} \right) 
           \Gamma_{ \mu } \Gamma_{ \dot{5} } 
       \right] \varepsilon_{i} $ 
and $ \nabla_{5} \varepsilon_{i} = \varepsilon^{\prime}_{i} $.}
\begin{align*}
0 &=   \langle \delta \Psi_{ \mu i } \rangle
   =   \partial_{ \mu } \varepsilon_{i}
     + \frac{1}{2} \left( \frac{ a^{\prime} }{a} \right)
       \Gamma_{ \mu } \Gamma_{ \dot{5} } \varepsilon_{i}
     - \frac{1}{2} 
       \left[- g_{R} \, \textrm{sgn}(x^{5}) \mathcal{W}_{crit} \right] 
       \Gamma_{ \mu } \left(iQ_{ij}\right) \varepsilon^{j}
\\
0 &=   \langle \delta \Psi_{5i} \rangle
   =   \varepsilon^{\prime}_{i}  
     - \frac{1}{2} 
       \left[- g_{R} \, \textrm{sgn}(x^{5}) \mathcal{W}_{crit} \right]
       \Gamma_{5} \left(iQ_{ij}\right) \varepsilon^{j}
\\ 
0 &=   \langle \delta \lambda^{ \tilde{x} }_{i} \rangle 
   = - \frac{i}{2} \Gamma^{5} 
       \phi^{\tilde{x} \prime}_{crit} \varepsilon_{i}
     + g W^{ \tilde{x} }_{crit} \, \varepsilon_{i}
     + \frac{i}{2} 
       \left[ 3 g_{R} \, \textrm{sgn}(x^{5}) 
              \mathcal{W}^{,\tilde{x}}_{crit} \right]
       \left(iQ_{ij}\right) \varepsilon^{j}.
\end{align*}
To solve these equations, split $ \varepsilon_{i} $ into 
$ \mathbb{Z}_{2} $ even $ ( \varepsilon^{+}_{i} ) $ and
$ \mathbb{Z}_{2} $ odd $ ( \varepsilon^{-}_{i} ) $ parts:
\begin{align}
\varepsilon_{i} &= \varepsilon^{+}_{i} + \varepsilon^{-}_{i}
\\
\varepsilon^{\pm}_{i} 
  &=   \frac{1}{2} 
       \left[
         \varepsilon_{i} \pm \Gamma_{ \dot{5} } 
         \left(iQ_{ij}\right) \varepsilon^{j}
       \right]
   = \pm \Gamma_{ \dot{5} } 
     \left(iQ_{ij}\right) \varepsilon^{ \pm j }.
\end{align}
Writing $ \langle \delta \Psi_{5i} \rangle = 0 $ in terms of
$ \varepsilon^{ \pm }_{i} $ and using
$ a^{\prime} / a 
  = - g_{R} \, \textrm{sgn}(x^{5}) \mathcal{W}_{crit} $ yields
\begin{equation}
\varepsilon^{ + \prime }_{i} + \varepsilon^{ - \prime }_{i}
  - \frac{1}{2} \left( \frac{ a^{\prime} }{a} \right)
    \left( \varepsilon^{+}_{i} - \varepsilon^{-}_{i} \right) = 0. 
\end{equation}
This determines the $ x^{5} $ dependence of 
$ \varepsilon^{\pm}_{i} $ to be
\begin{equation}
\varepsilon^{\pm}_{i} 
  = a^{ \pm \frac{1}{2} } \varepsilon^{\pm}_{i} (x^{\mu}).
\end{equation}
Writing $ \langle \delta \Psi_{ \mu i } \rangle = 0 $ in terms of
$ \varepsilon^{ \pm }_{i} $ and using
$ a^{\prime} / a
  = - g_{R} \, \textrm{sgn}(x^{5}) \mathcal{W}_{crit} $ yields
\begin{equation}  
\partial_{\mu} \varepsilon^{+}_{i} + \partial_{\mu} \varepsilon^{-}_{i}
  + \left( \frac{ a^{\prime} }{a} \right)
    \Gamma_{\mu} \Gamma_{ \dot{5} } \varepsilon^{-}_{i} = 0.   
\end{equation}
This determines that
\begin{equation}
\varepsilon^{+}_{i}(x^{\mu}) = \varepsilon^{+(0)}_{i},
\quad
\varepsilon^{-}_{i}(x^{\mu}) 
  = \left(
      1 - \frac{ a^{\prime} }{a} x^{\mu} \Gamma_{\mu} \Gamma_{ \dot{5} }      
    \right) \varepsilon^{-(0)}_{i}
\end{equation}
where 
$ \varepsilon^{\pm(0)}_{i} $ 
are constant (projected) spinors.
The Killing spinors are thus
\begin{equation}
\varepsilon_{i} 
  =   a^{ \frac{1}{2} } \varepsilon^{+(0)}_{i}
    + a^{-\frac{1}{2} } 
      \left(
        1 - \frac{ a^{\prime} }{a} x^{\mu} \Gamma_{\mu} \Gamma_{ \dot{5} }
      \right)
      \varepsilon^{-(0)}_{i}.
\end{equation}
This solution corresponds to $ \mathcal{N} =2 $ supersymmetry in the bulk.  
On the 3-branes (where the $ \mathbb{Z}_{2} $ odd fields vanish) there is 
$ \mathcal{N} = 1 $ supersymmetry.
No constraints on the Killing spinors arise from
$ \langle \delta \lambda^{ \tilde{x} }_{i} \rangle = 0 $.

\section{\label{Discussion}Discussion}

The model of Section \ref{Orbifold} is one example of the locally 
supersymmetric two brane Randall-Sundrum scenarios which can be 
constructed from
$ D \! = \! 5 $, $ \mathcal{N} \! = \! 2 $
gauged Yang-Mills/Einstein supergravity.
Here, the construction starts from a
$ D \! = \! 5 $, $ \mathcal{N} \! = \! 2 $
gauged Yang-Mills/Einstein supergravity theory
with scalar manifold
$ \mathcal{M} = SO(1,1) \times SO(2,1) / SO(2) $
and gauge group
$ U(1)_{R} \times SO(2) $.
Next, the $ U(1)_{R} $ gauge coupling $ g_{R} $ is replaced by 
$ g_{R} \, \textrm{sgn}(x^{5}) $ and the fifth dimension is compactified
on $ S^{1} / \mathbb{Z}_{2} $.
The conditions of local supersymmetry for the bulk plus brane system admit
the vacuum solution ($ \ref{Ansatz} $) and yield the relations
$ \mathcal{T}^{(0)} = - \mathcal{T}^{( \pi \rho )}
                    = - \Lambda / k = 6k $.
This vacuum preserves
$ \mathcal{N} \! = \! 2 $ supersymmetry in the $ AdS_{5} $ bulk and
$ \mathcal{N} \! = \! 1 $ supersymmetry on the Minkowski 3-branes.
It would be interesting to generalize this construction to the case in 
which the scalar manifold has the form described in Footnote \ref{Jordan}.
This is left to future work.

\section*{Acknowledgements}

I would like to thank Alon Faraggi, Zygmunt Lalak, and Francesco Toppan 
for discussions. 
I would also like to thank the Rudolf Peierls Centre for Theoretical 
Physics at the University of Oxford for hospitality during the initial 
stages of this project. 
This work is supported by FAPERJ.


\end{document}